\documentclass{appolb}
\usepackage{graphicx}
\usepackage[latin1]{inputenc}
\usepackage{pstricks,pst-node,pst-text,pst-3d}
\usepackage{amsmath}
\usepackage{hyperref}

\newcommand{\be}{\begin{equation}}
\newcommand{\ee}{\end{equation}}
\newcommand{\bea}{\begin{eqnarray}}
\newcommand{\eea}{\end{eqnarray}}

\newcommand{\ben}{\begin{eqnarray*}}
\newcommand{\een}{\end{eqnarray*}}

\begin{document}
\title{Singularities and cyclic universes%
\thanks{Presented at 3rd POTOR Conference', Krak\'ow - 28 September 2016}%
}
\author{Mariusz P. D\c{a}browski
\address{Institute of Physics, University of Szczecin, Wielkopolska 15, 70-451 Szczecin; National Centre for Nuclear Research, Andrzeja So{\l}tana 7, 05-400 Otwock, and Copernicus Center for Interdisciplinary Studies, S{\l}awkowska 17, 31-016 Krak{\'o}w.}
}

\maketitle
\begin{abstract}
The models of cyclic universes and cyclic multiverses based on the alternative gravity theories of varying constants are considered. 
\end{abstract}
 
\section{Introduction - singularities in cosmology.}

Cosmology is facing the problem of singularities -- places where general relativity fails and one asks if one is able to avoid them under some generic conditions. The best definition of singularities is by geodesic incompletness \cite{Wald} which allows to practically detect them without ``adopting'' them into the theory. This is a ``minimalistic'' approach which does not tell us the fuller and richer nature of these singularities, e.g. how do they influence physical and geometrical quantities we define. There are more subtleties of the matter and there are some tools to investigate them. 

For example, Tipler \cite{Tipler} says that a singularity is strong if the integral 
$I_j^i (\tau) = \int_0^{\tau} d\tau' \int_0^{\tau'} d\tau'' |R^i_{ajb}u^a u^b|$ diverges at $\tau = \tau_s$. Kr\'olak \cite{Krolak} says that a singularity is strong if another integral $I_j^i(\tau) = \int_0^{\tau} d\tau' |R^i_{ajb}u^a u^b|$ diverges at $\tau = \tau_s$. 
According to the above, various types of singularities within the framework of relativistic isotropic cosmology were found \cite{MPDCCP}.  Among them ($a$ - the scale factor, $\varrho$ - mass density, $p$ - pressure, $w$ - barotropic index) type 0 or standard Big-Bang (Big-Crunch) $a \to 0$, $p \to \infty$, $\varrho \to \infty$; Type I - Big-Rip $a(t_{s}) \to \infty$ ($t_{s} < \infty$), $p \to \infty$, $\varrho \to \infty$; Type II - Sudden Future $a(t_s) =$ const., $\varrho =$ const., $p \to \infty$; Type IIg - Generalized Sudden Future $a(t_s)$= const., $\varrho=$ const., $p=$const., $\dddot{a} \to \infty$ etc., $w < \infty$; Type III - Finite Scale Factor $a(t_s)$ = const., $\varrho \to \infty$, $p \to \infty$; Type IV - Big Separation: $a(t_s)$= const., $p=\varrho=0$, $w \to \infty$, $\dddot{a} \to \infty$ etc.; Type V - $w$-singularity $a(t_s)$= const., $p=\varrho=0$, $w \to \infty$; Little-Rip $a(t_{s}) \to \infty$, $\varrho(t_{s}) \to \infty$ ($t_{s} \to \infty$) and Pseudo-Rip $\varrho(t_{s}) < \infty$ ($t_{s} \to \infty$). Out of the above only types 0, I, Little-Rip are Tipler-strong while types 0, I, III and Little-Rip are Kr\'olak-strong. It is then clear that they are not the same and possess a property of being ''weaker'' or ''stronger''. 

\section{Varying constants against singularities.}

It has been shown \cite{Houndjo} that quantum effects may change the strength of exotic singularities (e.g.  SFS can be changed into FSF etc.). Similar effects can happen in the alternative gravity theories which describe variation of fundamental constants such as the speed of light $c$ (which have been applied to solve standard cosmology problems -- the horizon and the flatness problem \cite{Moffat}) and gravitational constant $G$ \cite{BM}. For such theories Einstein equations generalize into \cite{BM} (in the simplest minimally coupled and a preferred frame version)
\be 
\varrho(t) = \frac{3}{8\pi G(t)}
\left(\frac{\dot{a}^2}{a^2} + \frac{kc^2(t)}{a^2}
\right),\hspace{0.2cm}
p(t) = - \frac{c^2(t)}{8\pi G(t)} \left(2 \frac{\ddot{a}}{a} + \frac{\dot{a}^2}{a^2} + \frac{kc^2(t)}{a^2} \right),
\ee
and the energy-momentum ``conservation law'' read as
\be
\label{conser}
\dot{\varrho}(t) + 3 \frac{\dot{a}}{a} \left(\varrho(t) + \frac{p(t)}{c^2(t)} \right) = - \varrho(t) \frac{\dot{G}(t)}{G(t)}
+ 3 \frac{kc(t)\dot{c}(t)}{4\pi Ga^2}~.
\ee

In Ref. \cite{JCAP13} we have presented series of examples in which we have proved that one is able to  
regularize an SFS singularity by varying $c$ provided that light eventually stops moving. Physical model for that is given in loop quantum cosmology (LQC) in the anti-newtonian limit $c = c_0 \sqrt{1- \varrho/\varrho_c} \to 0$ for $\varrho \to \varrho_c$, with $\varrho_c$ being the critical density \cite{Cailettau}. The low-energy limit $\varrho \ll \varrho_0$ gives $c \to c_0$. 

Similarly, it is possible to regularize singularities by varying $G$ provided it is in the strong coupling limit $G \to \infty$. 
This mechanism have been applied in cyclic brane motivated scenarios \cite{Khoury} where instead of $G \to \infty$ one has 
some special coupling of a scalar field $\beta (\phi)$ of gravity in the Lagrangian of a 4-dimensional theory in the Einstein frame. 

\section{Cyclic universes and the multiverse. }

In Ref. \cite{MNRAS} we have considered a positive curvature $(k=+1)$ and $\dot{c}=0$ sinusoidal pulse Friedmann model with the scale factor and gravitational constant evolution
\be
a(t)= a_{0} \left| \sin \left( \pi \frac{t}{t_{c}} \right) \right| , \hspace{0.5cm} G \left( t \right)= \frac{G_{0}}{a^2(t)}, 
\ee
($a_0, G_0, t_c =$ const.). We have also investigated the tangential pulse model with 
\be
a(t)= a_{0} \left| \tan \left( \pi \frac{t}{t_{s}} \right) \right|~,\hspace{0.5cm}G \left( t \right) = \frac{4G_s}{\sin^2{\left( 2 \pi \frac{t}{t_s} \right)}} ~.
\ee
In both cases we have shown that the mass density and pressure are nonsingular, though the scale factors can either be zero or infinity. 

What is interesting for the varying $G$ models here is that the strong energy condition is fulfilled so that gravity keeps being attractive, but in our case its attractivity is overbalanced by the strong coupling limit $G \to \infty$ at the Big-Bang singularity. The mechanism here is similar to a cyclic brane universe \cite{Khoury}, where the 5th dimension (orbifold) collapses (when boundary branes collide), while the 4-dimensional theory has no singularity at all. The role of the coupling $\beta(\phi) \propto 1/a$ there is played by the running gravitational constant $G(t) \propto 1/a^2(t)$ here, which regularises the mass density and pressure and one has a kind of a ''singular bounce'' in the scale factor $a(t)$, and a ''non-singular'' bounce in the mass density and pressure.

Using the generalized 2nd law of thermodynamics of varying $c$ and varying $G$ universes for total non-decreasing entropy (multiverse obeying 2nd law of thermodynamics, but not individual universes) we have attempted to create a mock model of a cyclic multiverse (the doubleverse) for which the two scale factors are equal and sinusoidal \cite{MNRAS}
$ a(t) =  a_1 \left( t \right) =  a_2 \left( t \right) = a_0 \left| \sin{\left( \pi t/t_s\right)} \right|$, 
so that the geometrical evolution of the ''parallel'' universes is the same. However, this is not the case for the evolution of the physical constants $c$ and $G$ which is different in each universe. It is interesting that because of the same evolution, one can consider that the universe 1 {\it may replace its evolution} along the trajectory of the universe 2 {\it at the maximum expansion point} by quantum effects. It is a new option for the evolution of the universe -- now put in the context of the multiverse -- where some effects take place at the turning point. Similar phenomena (though considered to be appropriate to the same universe) were studied long time ago in the context of quantum cosmology \cite{PRD95}  and more recently developed under the name of the simple harmonic universe (SHU) scenario \cite{Vilenkin}. All this is in agreement with the claim that macroscopic quantum effects in cosmology are possible \cite{Kiefer}. In a more recent contribution \cite{Entangl} it has been shown that within the framework of third quantization picture, where one defines creation and annihilation operators as in quantum field theory one is able to consider parallel evolving universes as a pair of spontaneously created from the vacuum state remaining quantum mechanically entangled during their classical evolution. An interesting effect is that for such a pair the entropy of entanglement is also large at the maximum point of expansion presumably signalling strong quantum effects there. This happens apart from large entropy and temperature of entanglements at the points of classical singularities. 

\section{Conclusions.}

Curiously, one is able to differentiate quite a number of cosmological singularities leading to a blow-up of physical quantities such as scale factor, energy density, pressure, physical fields etc., and not to geodesic incompleteness. These singularities can be influenced by in alternative gravity theories with varying constants. Using these theories it is possible to create various {\it cyclic universe scenarios} and extend them into the {\it cyclic multiverse scenarios} with different evolution of the coupling constants and same geometries obeying the total 2nd law of thermodynamics. These universes are classically disconnected, but they can be quantum mechanically entangled and the effect of entanglement can be detected in individual universes as the temperature or the entropy of entanglement perhaps signalling at the cosmic microwave background or the large-scale structure of the universe. 


This paper was financed by the Polish National Science Center Grant DEC-2012/06/A/ST2/00395.

\end{document}